\documentclass[mathleft
]{an}
\usepackage{graphicx}
\usepackage{times}
\begin{document}

% The following seven commands are intended for editorial usage and should be ignored by
% the author(s).
\Pagespan{789}{}% Document's page range. 
% If second parameter is left empty, the last page is computed automatically.
\Yearpublication{2010}%
\Yearsubmission{2010}%
\Month{11}%   
\Volume{999}%  
\Issue{88}% 
% \DOI{This.is/not.aDOI}% 

\title{An automatic pipeline analysing solar-like oscillating targets tested on CoRoT and simulated data}

\author{S. Mathur\inst{1}\fnmsep\thanks{Corresponding author:
  \email{savita@ucar.edu}\newline}
%Example 
%for footnote, note the usage of the \texttt{fnmsep}
%command as separator between institute number and footnote mark} 
\and  R.~A. Garc\'ia\inst{2}
\and C. R\'egulo\inst{3,4}
\and O.~L. Creevey\inst{3,4}
\and J. Ballot\inst{5}
\and D. Salabert\inst{3,4}
}
\titlerunning{An automatic pipeline analysing solar-like oscillating targets}
\authorrunning{Mathur et al.}
\institute{
High Altitude Observatory, NCAR, P.O. Box 3000, Boulder, CO 80307, USA
\and 
Laboratoire AIM, CEA/DSM -- CNRS - Universit\'e Paris Diderot -- IRFU/SAp, 91191 Gif-sur-Yvette Cedex, France
\and 
Universidad de La Laguna, Dpto de Astrof\'isica, 38206, Tenerife, Spain
\and  Instituto de Astrof\'\i sica de Canarias, 38205, La Laguna, Tenerife, Spain
\and  Laboratoire d'Astrophysique de Toulouse-Tarbes, Universit\'e de Toulouse, CNRS, F-31400, Toulouse, France}

\received{}
\accepted{}
\publonline{later}


\abstract{%
 The launch of the Kepler mission on 7th March 2009 opened a new bright future for the search of extra-solar planets while a huge amount of stars will be observed leading to the opportunity to better understand stellar evolution. This will allow us to probe different regions in the HR diagram and put more constraints on the stellar models. Up to now the asteroseismic missions such as MOST and CoRoT were providing some solar-like stars at a very slow cadence. But to study the several hundreds of solar-like oscillating stars that will be observed during the Kepler survey phase, an analysis devoted to one star is impossible if we want to have as much information as we can in a small period of time. We describe here our pipeline, A2Z, which calculates the global parameters of the stars (rotation period, mean large spacing of the acoustic modes, maximum amplitude of the modes), fits the modes globally, and estimates the radius and mass of the stars. This pipeline has been tested on simulated stars and applied to real data from CoRoT.
}

\maketitle

\section{Introduction}

The study of stellar evolution is going through an important revolution. Helioseismology had already allowed us to improve our knowledge on the solar interior, such as density, rotation, position of the base of the convection zone,and  He content, showing the importance of this useful tool. 

During the last decade, the development of asteroseismology started in many ways: through ground-based observations campaigns and thanks to many space missions allowing the observations of solar-like oscillations over a longer period of time (WIRE, MOST, CoRoT).

More recently, the launch of the NASA mission, Kepler, dedicated to the search for exoplanets has also extended the study of stellar oscillations to a few hundreds of solar-like oscillating stars.

To face the arrival of a huge amount of data, several teams have developed different pipelines (Huber et al, 2009; Hekker et al. 2010) in the framework of the AsteroFLAG group. We will describe the A2Z pipeline (Mathur et al. 2010a). This pipeline is divided into several packages: the search for the frequency-range of the p-mode bump by looking for the large spacing (average distance in frequency between consecutive nodes for the same mode degree), the fitting of the background, the estimation of the maximum amplitude per radial mode as well as the central maximum frequency. At this point, we can try to infer the radius and the mass of the stars from all this information (e.~g. Stello et al. 2009). Then, if the signal-to-noise ratio is high enough we can obtain the characteristics of the p modes through a global fitting on the power spectrum. For the first package, we also have several methods that enables us to cross-check the results and to attribute a confidence level. After describing the methodology used in the pipeline, we will apply it to the Sun and to a few CoRoT targets.

\section{Pipeline packages}

The \emph{A2Z} pipeline is constituted of 3 parts: the first one for retrieving the global parameters of the acoustic modes as well as a first estimation of the mass and radius using the scaling laws, the second one to fit the modes globally, and the last one to estimate the radius and mass of the stars by using stellar modelling. We will focus here on the determination of the global parameters of the acoustic modes, the estimation of the guesses, and the inference of the radius of a star.

If the signal-to-noise ratio is high enough the global-fitting code can be run and it will infer the central frequency ($\nu_i$), the amplitude ($A_i$), the width ($\Gamma_i$), and the rotational splittings ($\delta \nu_i$) of each mode.

To compute the Power Spectrum Density (PSD), we use the Lomb-Scargle algorithm, which takes into account the irregularity of the sampling rate of the time-series.

In the case that there are gaps in the time series, we use the in-painting algorithm described by Sato et al. (2010).
For very low signal-to noise ratio data, we also plan to apply the curvelets (Garc\'ia et al. 2010).

\begin{figure*}
\centering
\includegraphics[width=150mm,height=110mm]{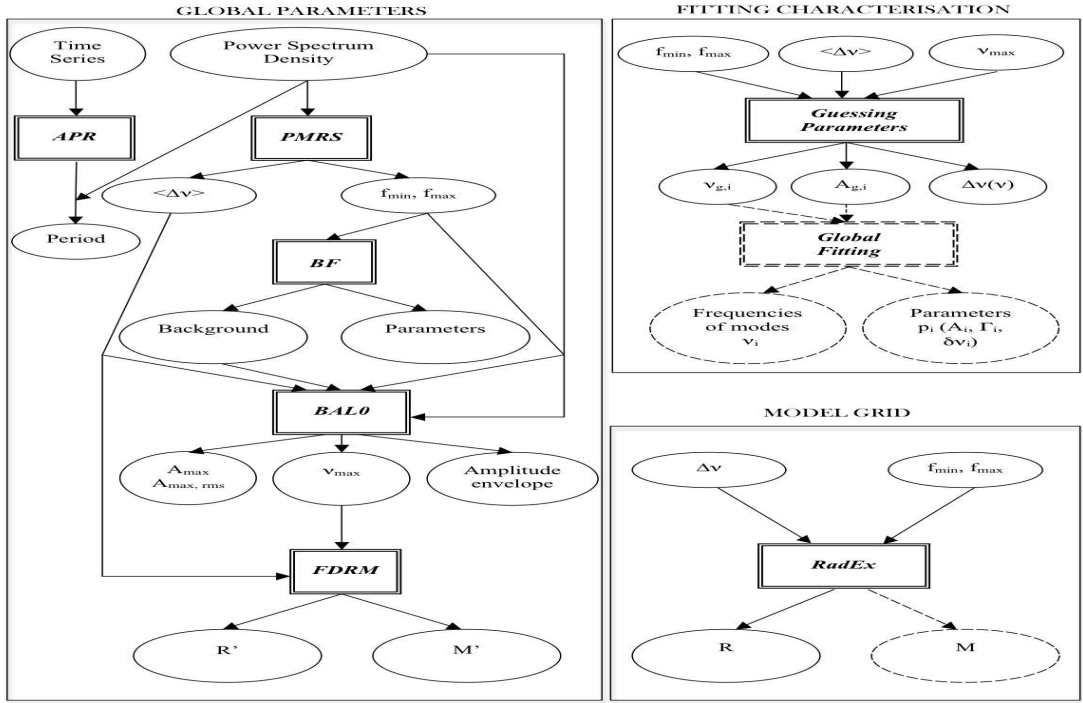}
\caption{Workflow diagram of the \emph{A2Z} pipeline. The circles are the inputs and outputs of the different packages (boxes). Each output parameter is given with its uncertainty (extracted from Mathur et al. 2010a).}
\label{workflow}
\end{figure*}

\subsection{Average Photospheric Rotation}

Some information on the surface rotation of the star can give some constraints on the rotational splitting and the inclination while fitting the modes. Indeed, it is very difficult to estimate these two values only with the fitting procedure as they are mostly correlated  because of the short lifetime of the modes (Ballot et al. 2006, 2008)

\begin{figure}
\includegraphics[width=50mm,height=76mm, angle=90]{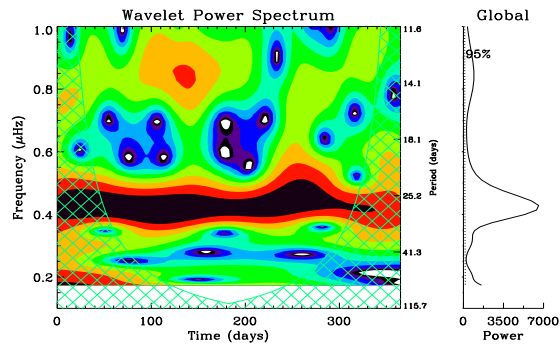}
\caption{ Wavelet power spectrum for 1 month of VIRGO data (left). Global power spectrum (right)
(from Mathur et al. 2010).}
\label{wave_sun}
\end{figure}

We apply two different methods to retrieve the rotation period of the stellar surface. First we can compute the PSD and look at the highest peak at low frequency, below 5~$\mu Hz$. However, sometimes, this method does not enable us to distinguish between the fundamental period and its first harmonic. Thus we also use a time-frequency analysis, with the wavelet tools. We use the Morlet wavelet to calculate the correlation between this wavelet that we slide along the time-series. The frequency of the wavelet takes many values in a given range.
Fig.~\ref{wave_sun} shows the results, wavelet power spectrum (WPS) for 1 month of solar data taken by the VIRGO (Frohlich et al. 1995) instrument aboard SoHO. If we collapse the WPS along time, we obtain the global wave\-let power spectrum (GWPS) on the right panel of Fig.~\ref{wave_sun}. We see on Fig.~\ref{wave_sun}  a large dark pattern showing that there is some continuous power along time around a period of 27 days as well as a large peak in the GWPS. We can also provide a confidence level (95\% is the value chose in our study). Thus we can give an estimation of the rotation period of the solar surface with more than 95\% confidence level.

\subsection{P-Mode Range Search}

We are looking for the frequency range of the p-mode excess power as well as an estimation of the mean large spacing, $\langle \Delta \nu \rangle$. We use two different methods to cross-check our results and to select only those for which we are confident.

\begin{figure}[h!]
\includegraphics[width=50mm,height=76mm, angle=90]{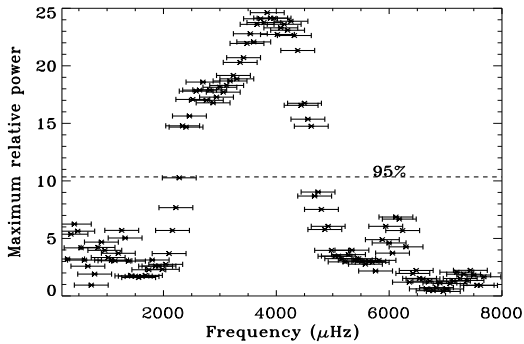}
\caption{ Frequency-range of the p modes for 1 month of VIRGO data.}
\label{frange_sun}
\end{figure}

\noindent Method 1: we estimate$\langle \Delta \nu \rangle$ by calculating the Power Spectrum (PS) of PS (from 200 to 6000 µHz). We compute the PS of PS (or PS2) on sub-series of 600~$\mu$Hz. We obtain the frequency-range of the p modes, $f_{min}$ and $f_{max}$ with a given confidence level: from 85 to 95\% (level 2), from 70 to 85\% (level 1), and below 70\% (level 0) (see Fig.~\ref{frange_sun}). 

\noindent Method 2: PSSPS (Power Spectrum of a Short or Small part of the Power Spectrum). Here, we estimate $\langle \Delta \nu \rangle$ from the standard stellar parameters. Then, we search for the spacing in the PS of a short slice of the PS where the p-mode excess power appears (R\'egulo \& Roca Cort\'es 2002).

\noindent We have a 3-step strategy, which is the following:
\begin{itemize}
\item Run method 1 and 2;
\item Run method 2 with the results of method 1 as inputs;
\item Run method 1 with the results of method 2 as inputs.
\end{itemize}
To select the most reliable results, we dismiss the step in which we obtain the highest number of common stars with level of 0. Among the two remaining steps in which we remove the stars with the confidence level of 0, we choose the step where we have the highest number of coincidences. 

\subsection{Background Fitting}

For the background fit, we use a fit of three components, one or several Harvey laws (Harvey 1986), one power law for the activity effects, and a flat photon noise, W:

\begin{equation}\label{eq:bgmodel}
 B(\nu)=W+\frac{A}{1+\left(2\nu/\nu_c\right)^\alpha}+a\nu^{-b},
\end{equation}

\noindent where we have six free parameters. The PSD is fitted without the mode region and with a maximum likelihood estimator.

\subsection{Bolometric Amplitude of the $\ell$=0 mode}

We look for the amplitude envelope of the PS. First, we smooth the PSD corrected from background over 1 or 2  $\times \langle \Delta \nu \rangle$. We fit the PSD with a Gaussian to get the maximum power $P_{max}$ and $\nu_{max}$. We convert the power into the amplitude per radial mode $A_{max}$  (Kjeldsen et al. 2008) and to bolometric amplitude (Michel et al. 2009).

\subsection{First Determination of the Radius and the Mass}

Based on the scaling laws (Kjeldsen \& Bedding 1995), we can calculate the radius and the mass of the star with the following formulae:

\begin{equation}
\frac{R'}{R_\odot}=\Big( \frac{135}{\langle \Delta \nu \rangle}\Big)^2 \Big(\frac{\nu_{\rm max}}{3050} \Big) \Big(\frac{T_{\rm eff}}{5777} \Big)^{1/2}\\
\end{equation}

\begin{equation}
\frac{M'}{M_\odot}=\Big(\frac{135}{\langle \Delta \nu \rangle} \Big)^4 \Big(\frac{\nu_{\rm max}}{3050} \Big)^3 \Big(\frac{T_{\rm eff}}{5777} \Big)^{3/2}. \\
\end{equation}

\subsection{Guessing Parameters}

We determine of the table of guesses of individual mode frequencies and variation of $\langle \Delta \nu \rangle$ with frequency based on the results obtained with PMRS. We look for the highest peaks in the range [$f_{\rm min}$, $f_{\rm max}$] and that are approximately separated by $\langle \Delta \nu \rangle$, allowing an uncertainty of 15\%.

\subsection{Radius Extractor}

Here, we use a pre-calculated grid of evolutionary models to obtain an initial guess of the parameters such as mass and age.  With a minimization algorithm, we obtain a better estimation of the radius and mass.

\section{AsteroFLAG/Aarhus simulated stars}

\begin{figure}[h!]
\includegraphics[width=76mm,height=50mm]{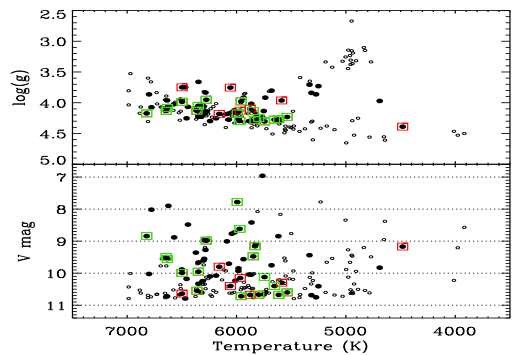}
\caption{ Distribution of the correct results obtained with the simulated stars from the Aarhus-asteroFLAG simulator. Full black circles represent the stars for which the correct value of the large spacing was obtained. The squares are the stars selected to determine the radius: the converging models (green) and the diverging models (red).
}
\label{dist_res}
\end{figure} 

\begin{figure}
\includegraphics[width=76mm,height=50mm]{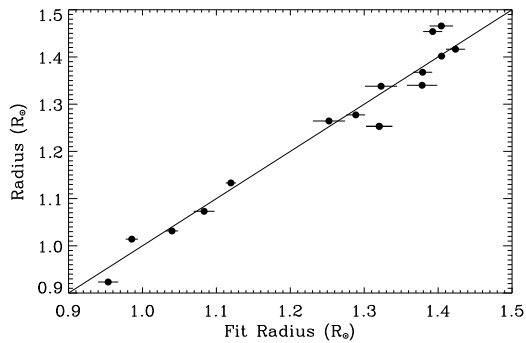}
\caption{The true radius versus the fitted input radius for a selection of the simulations.  The continuous line is the x=y line.
}
\label{radii_est}
\end{figure}

\noindent The A2Z pipeline was tested on 176 stars of magnitude from 7 to 11 that had been simulated by the Aarhus-AsteroFLAG simulator (Chaplin et al. 2008). These time-series are one-month long to simulate the data obtained with the Kepler mission. Following the 3-step strategy, we obtain a common mean large spacing between the 2 methods of the package PMRS for 71 stars, among which 67 are the correct ones.
We have applied the other packages BF and BAL0 to calculate the background fit and the maximum amplitude per radial mode for these 67 stars. The amplitudes are often underestimated.
Finally, the RadEx package was used on a few tens of stars to estimate their radii. Fig.~\ref{radii_est} shows the comparison between the radius input and the result of the fit for a few tens of stars.

\section{CoRoT solar-like stars}

\begin{figure}
\includegraphics[width=50mm,height=76mm, angle=90]{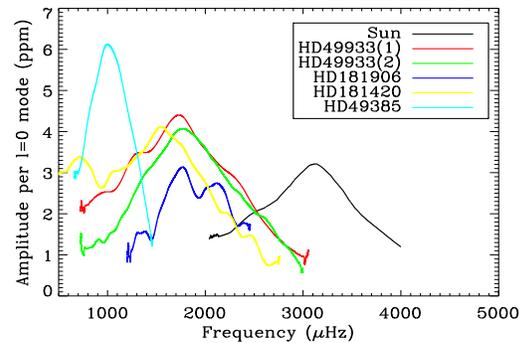}
\caption{Smoothed bolometric amplitude per radial mode (rms) for the Sun, HD49933 (run 1 and 2), HD181906, HD181420, and HD49385. }
\label{ampl_env}
\end{figure}

The A2Z pipeline has been applied to solar data from VIR\-GO, to several CoRoT targets such as, HD49933 (Appourchaux et al, 2008; Benomar et al. 2009), HD181906 (Garc\'ia et al. 2009), HD181420 (Barban et al. 2009), HD175726 (Mosser et al. 2009), HD49385 (Deheu\-vels et al. 2010), and HD~170987 (Mathur et al. 2010b), and to the Kepler data, solar-like oscillating stars and red giants (Bedding et al 2010; Chaplin et al, 2010; Stello et al. 2010). We give a few reseults in table 1. Fig.~\ref{ampl_env} shows the bolometric amplitude envelope per radial mode for the Sun and the CoRoT targets.

\begin{table}
% \centering%%%
\caption{Results of the A2Z pipeline on real data}
\label{tlab}
\begin{tabular}{cccccc}\hline
Star & Period & $\langle \Delta \nu \rangle$& $\epsilon$ & $R$ & $\epsilon$\\ 
 & (days) & ($\mu$Hz) &  ($\mu$Hz) & ($R_{\odot}$) & ($R_{\odot}$)\\
\hline
Sun & 27.8 & 135.46 & 3.26 & 1.016 & 0.023 \\
HD49933 & 3.3 & 86.14 & 1.58 & 1.398 & 0.04\\
HD181906 & 2.8 & 86.53 & 1.69 & 1.613 & 0.032\\
HD181420 & 2.6 & 75.35 & 1.53 & 1.42 & 0.018\\
\hline
\end{tabular}
\end{table}

%\section{Conclusions}

\acknowledgements
This work has been partially supported by: the CNES/GOLF grant at the Service d'Astrophysique (CEA/Sa\-clay) and the grant PNAyA2007-62650 from the Spanish National Research Plan. This work benefited from the support of the International Space Science
Institute (ISSI), through a workshop programme award. It was also partly
supported by the European He\-lio- and Asteroseismology Network (HELAS), a
major international collaboration funded by the European Commission's
Sixth Framework Programme. SoHO  is a space mission of international cooperation between ESA and NASA. CoRoT (Convection, Rotation, and planetary Transits) is a mini-satellite developed by the French Space agency CNES in collaboration with the Science Programs of ESA, Austria, Belgium, Brazil, Germany, and Spain (Mi\-chel et al. 2008).

%\newpage%%%%%%%%%%%%%%%%%%%%%%%%%%%%%%%%%%%%%%%%%%%%%%%%%%%%%%

\appendix

%\section{This is the title of the first appendix}
%Larger tables, collections of images, spectra or similar kind of data shall be 
%presented in the appendix section rather than in the main body of the 
%text. Several appendices can be separated by the \verb+\section{+{\it title
%of appendix}\verb+}+ command. They are enclosed in the 
%\verb+appendix+ environment.

\end{document}